\documentclass[a4paper,12pt]{article}
\begin{document}
\def\baselinestretch{1.5}
{\def\title{The equilibration of a hard-disks system}
\def\author{V.M. Somsikov}
\def\correspondence{Prof. V.M. Somsikov}
\def\correspondenceshort{Prof. Somsikov}
\def\date{\today}
\def\addra{Laboratory of Physics of the geogeliocosmic relation}
\def\addrb{Institute of Ionosphere}
\def\addrc{Kamenskoe Plato}
\def\addrd{Almaty, 480020, Kazakhstan}
\def\tel{~~  ~~~+8-3272-548074~~~~~~~~~~~~~~~~~~~~~}
\def\fax{~~  ~~~+8-3272-658085~~~~~~~~~~~~~~~~~~~~~}
\def\email{~~  ~~ nes@kaznet.kz ~~~~~~~~~~}}
\small
\title {Equilibration of a hard-disks
system}
\author{V.M. Somsikov }
\date{\it{Laboratory of Physics of the geogeliocosmic relation,\\
Institute of Ionosphere, Almaty, 480020, Kazakhstan}} \maketitle
\begin{abstract}
The process of equilibration of a  colliding hard-disks system is
studied in the framework of classical mechanic. The method
consists of dividing the nonequilibrium system into the
interacting subsystems; the evolution one of these subsystems is
analyzed employing generalized Lagrange and Liouville equations.
The subsystem-subsystem interaction force is considered as an
evolution parameter. The mechanism by which its system
equilibrates is described.\\

\end{abstract}
\bigskip
\begin {center}
{1. Introduction}
\end {center}

The existing methods of analyzing many-body systems based on the
models of disks and spheres have been shown to be highly
effective, see e.g. [Kryilov, 1950; Sinai, 1970; Beijeren {\&}
Dorfman, 1995]. Nevertheless, it is not clear whether it is
possible to prove rigorously, within the framework of classical
mechanics laws, that equilibrium is established in a sum system,
see [Petrosky {\&} Prigogine, 1997; Zaslavsky, 1984,1999]. Such an
enterprise would demand new methods and new approaches.

Some features of the nonequilibrium evolution of colliding disks
were studied and reported in [Somsikov, 1996, 1998, 2001a]. This
approach enables one to carry out analytical and numerical study
of the evolution process within the framework of classical
mechanics on the basis of equation of motion of hard disks. These
equations are written down for a matrix of colliding disks. It was
found that the establishment of equilibrium was determined by the
rate of decrease of forces between disks group, [Somsikov,
1996,2001b].

In the present work the process of establishment of equilibrium
state for a hard-disks system within the framework of the laws of
the classical mechanics is studied. The principal feature of this
study is that the interaction force of subsystems into which a
disks-system is divided, is used as evolution parameter.

This allows us to avoid application of probabilistic concepts for
an explanation of the mechanism of equilibration. Moreover it helps us
understand the nature of the probabilistic laws in a systems of the
classical mechanics.

The work is constructed as follows. Using D'Alambert principle and
equations of motion for disks, we construct generalized Lagrange,
Hamiltonian and Liouville equations for a subsystem, selected from
the full system. With the help of these
equations, we analyze peculiarities of evolution of the
disks-system  from a nonequilibrium state to equilibrium.

The study is based on the following method. First we prepare a
nonequilibrium disks system. A macroscopic subsystem of disks is
selected from this system. The evolution of forces on this
subsystem by other subsystems is analyzed.
\begin {center}
{2. Generalized  equations of the classical mechanics}
\end {center}

It is known that the methods, based on canonical Lagrange or
Hamilton equations, can be used successfully for investigating
many-body systems, which is not faraway from an equilibrium state
see [Lanczos, 1962; Landau, 1976; Landau {\&} Lifshits, 1974]. At
the same time these methods have inherent problem when we attempt
to use them for the study of strongly nonequilibrium systems,
especially ones with nonholonomic connections and polygenic forces
(see [Lanczos, 1962]). But in the real world, the majority of
systems are of this type. Indeed, even a disks-systems in the
space with ideally reflecting walls are nonholonomic [Goldstein,
1975]. Therefore for considering the process of evolution to
equilibrium state in a hard-disks system, we should modify the
equations of the classical mechanic so that they are applicable to
nonequilibrium systems. For deriving these equations, we must take
into account the fact that the forces acting between subsystems,
in general, cannot be expressed through a potential function; the
work done in going from one point to the other in a configuration
space depends on the path taken. These equations will allow us to
consider process equilibration, by using the interaction force of
subsystems, as a dynamic evolutionary parameter. In accord with
Lanczos [1962], we shall name these as "generalized equations".

Let us take a system, which consist of $N$ disks. Divide this
system into $R$ subsystems, so that in each subsystem will be of
$T$ disks. Therefore, $N=RT$. The energy of the system is
constant. It is equal to the sum of internal energies of all
subsystems and interaction energies between subsystems. Let us
selected one of them, which we call $p$-subsystem. Let
$\delta{W}_{a}^{p}$, be the virtual work of active forces in
$p$-subsystem. In the general case, this work can be express as
follows:
$\delta{W}_{a}^{p}=\sum\limits_{k=1}^{T}\sum\limits_{s=1}^{N-T}F_{ks}^p\delta{r_k}=
\sum\limits_{k=1}^{T} F_{k}^p\delta{r_k}$, where $k=1,2,3...T$ are
disks of $p$ - subsystem, $s=1,2,3...N-T$ are external disks that
interacts with the $p$ -disks. $F_{ks}^p$ is the interaction force
of $k$ - and $s$ - disks, $\delta{r_k}$ is the virtual
displacement of the $k$ -disk,
$F_{k}^p=\sum\limits_{s=1}^{N-T}F_{ks}^p$. The virtual work of the
interaction force of internal disks of $p$-subsystem is equal to
zero.

In case of the pair interaction the virtual work of external
forces is of the form: $\delta{W}_{a}^{p}= \sum\limits_{k=1}^{T}
F_{k}^p\delta{r_k}=\sum\limits_{k=1}^{T}F_{ks}^p\delta{r_k}$,
because in this case $F_{k}^p = F_{ks}^p$. The inertial force can
be presented as
$\delta{W}_{in}^p=\sum\limits_{k=1}^T\dot{V}_k\delta{r_k}$. The
sum of the active and the inertial forces constitute as effective
forces. The principle of D'Alambert asserts that the work of
effective forces is always zero [Lanczos, 1962], i.e.
\begin{equation}
{\delta\overline{W}_q^p=\delta{W}_{in}^p-\delta{W}_a^p}.\label{eqn1}
\end{equation}
The feature  of the virtual work above implies, that in general it
does not get reduced to complete differential.

From the equations of motion for hard disks see [Somsikov, 2001b],
it follows:
$\sum\limits_{p=1}^{R}(\sum\limits_{k=1}^{T}F_{ks}^p)\delta{r_k}=0$.
Therefore total active and inertial work for all subsystems at any
moment of time is equal to zero, i.e.
$\sum\limits_{p=1}^R{\delta{W}_{in}^p}=\sum\limits_{p=1}^R{\delta{W}_a^p}=0$.
This equality can take place in two cases: when the sum of nonzero
terms is equal to zero, or when each term of the sum is equal to
zero. Obviously the second case, appropriate for an equilibrium
state, takes place when $T\rightarrow\infty$. For this case, with
the help of the equations of motion for hard disks [Somsikov,
2001b], it is possible to record:
\begin{equation}
\sum\limits_{k=1}^T\varphi_{ks}\delta
(\psi_{ks}(t))\Delta_{ks}(t)|\Delta_{ks}(t)|=
\sum\limits_{k=1}^T{F_{ks}^p}=\sum\limits_{k=1}^T\dot{V_k}=0,
\label{eqn2}
\end{equation}
where $V_k$ is a $k$-disks velocity; $\varphi_{ks}=i\beta
e^{i\vartheta_{ks}}$; $i$ is an imaginary unit;
$\beta=sin\vartheta_{ks}$; $\vartheta_{ks}$ is scattering angle
for $k$ and $s$ colliding disks, which varies from $0$ to $\pi$;
$\delta(x)$ is a delta function; $\psi_{ks}=1-|l_{ks}|$;
$l_{ks}(t)=r_{ks}^0+\int \limits_{0}^{t}\Delta_{ks}dt$ are
distances between centers of colliding disks;
$r_{ks}^0=r_{k}^0-r_{s}^0$ are initial values of disks
coordinates; $\Delta_{ks}=V_k-V_s$ are relative velocities. The
impact parameter $d_{ks}=cos\vartheta_{ks}$ is determined by the
distance between centers of the colliding disks in the complex
plane with real axis $x$ and imaginary axis $y$. The value of the
impact parameter is:
$d_{ks}=Im(l_{ks}\Delta_{ks})/|l_{ks}\Delta_{ks}|$. The $k$-disk
swoops on the $j$- disk lying along the $x$ - axis. Mass and
diameter of each disk is set equal to 1. Boundary conditions are
either periodical or hard walls. Equating the right-hand side of
Eq. (2) to zero implies, that the selected $p$-subsystem is in a
stationary state.

To derive the general Lagrange equation for $p$-subsystem,
let us transform D'Alambert equation (1) by multiplying it by $dt$,
and integrating it over an interval from $t_1$ to $t_2$. In general
we have,

$\int\limits_{t_1}^{t_2}{\delta{\bar{W_q^p}}}dt=
\int\limits_{t_1}^{t_2}{\sum\limits_{k=1}^T[{\frac{d}{dt}}V_k-
\sum\limits_{j\neq k}^T{F_{kj}^p}-F_{k}^p]\delta{r_k}dt}$ =

$\delta\int\limits_{t_1}^{t_2}
{\frac{1}{2}}\sum\limits_{k=1}^T{V_k^2}dt-
\int\limits_{t_1}^{t_2}[\sum\limits_{k=1}^T(F_k^p+
\sum\limits_{j\neq k}^T F_{jk}^p)\delta{r_k}]dt-
{[\sum\limits_{k=1}^T{V_k}{\delta{r_k}}]}|_{t_1}^{t_2}$ (2a)

In Eq. (2a) the term $\sum\limits_{j\neq k}^T{F_{kj}^p}$
determines the force of interaction in the $p$-subsystem, $k$ and
$j$ are the colliding disks of the $p$-subsystem. The term
${\sum\limits_{k=1}^T}F_k^p$ is the force on the $p$-subsystem
from the rest. On demanded at the ends of the interval $[t_1,
t_2]$, the virtual displacements is zero. Then the last term in
(2a) will be zero.

Let us assume that a subsystem has equilibrated. Then for internal
forces of interaction in a subsystem we can set a function
$U(r_1,r_2,...r_T)$, for which the following condition is
satisfied:
$\int\limits_{t_1}^{t_2}[\sum\limits_{k=1}^T{\sum\limits_{j\neq
k}^T{F_{kj}^p}}\delta{r_k}]dt=
-\delta\int\limits_{t_1}^{t_2}U(r_1,r_2,...r_T)dt$. Here
$r_1,r_2...r_T$ are the coordinates of the $p$-subsystem disks. In
the general case it is impossible to express forces on
$p$-subsystem, as a gradient see [Lanczos, 1962]. In this case the
Eq. (2a), can be written as
\begin{equation}
\int\limits_{t_1}^{t_2}\delta{\bar{W_q^p}dt}=
\int\limits_{t_1}^{t_2}
[\sum\limits_{k=1}^T({\frac{d}{dt}}{\frac{\partial{L_p}}
{\partial{V_k}}}-{\frac{\partial{L_p}}
{\partial{r_k}}}-{F_{k}^p}){\delta{r_k}}]{dt}=0
\label{eqn3}
\end{equation}
In Eq. (3) we denote
$L_p=\sum\limits_{k=1}^T{\frac{V_k^2}{2}}+U(r_1,r_2,..r_T)$. If
the interaction of disks is potential, then $L_p$ will include
also internal potential energy of the $p$-subsystem -
$U(r_1,r_2,...r_T)$. Since any variations of integral in equation
(3) will be zero,we can set:
\begin{equation}
\sum\limits_{k=1}^T(\frac{d}{dt}\frac{\partial{L_p}}
{\partial{V_k}}-\frac{\partial{L_p}}{\partial{r_k}})=\sum\limits_{k=1}^T{F_k^p}=F_p
\label{eqn4}
\end{equation}
In the above $\sum\limits_{k=1}^T{F_k^p}=F_p$.

Eq. (4) is a generalized equation of Lagrange for a $p$-subsystem.
$F_p$ is the polygenic force acting on the $p$-subsystem. When
$F_{p}=0$, Eq. (4) transforms to a canonical equation of Lagrange
for equilibrium, conservative system. The equality, $F_{p}=0$, is
a sufficient condition for a stationary state.

Let us now derive Hamilton equation for the chosen $p$-subsystem.
The differential for $L_p$ can be written as,
$dL_p=\sum\limits_{k=1}^T(\frac
{\partial{L_p}}{\partial{r_k}}dr_k+\frac{\partial{L_p}}
{\partial{V_k}} d{V_k}) +\frac{\partial{L_p}}{\partial{t}}dt$,
where $\frac{\partial{L_p}}{\partial{V_k}}=p_k$ is disks momentum.
With the help of Lagrange transformation, it is possible to write:
$d[\sum\limits_{k=1}^Tp_k{V_k}-L_p]=
\sum\limits_{k=1}^T(-\frac{\partial{L_p}}{\partial{r_k}}dr_k+
{V_k}dp_k)-\frac{\partial{L_p}}{\partial{t}}dt$. Since
$\frac{\partial{H_p}}{\partial{t}}=-\frac{\partial{L_p}}{\partial{t}}$,
where $H_p=\sum\limits_{k=1}^Tp_k{V_k}-L_p$, we have from (4),
\begin{equation}
{\frac{\partial{H_p}}{\partial{r_k}}=-\dot{p_k}+F_{k}^p}.
\label{eqn5}
\end{equation}
\begin{equation}
{\frac{\partial{H_p}}{\partial{p_k}}={V_k}}.\label{eqn6}
\end{equation}
The above is a general Hamilton equations for the selected
$p$-subsystem. The right-hand side of Eq. (5) denote the external
forces, which act on $p$-subsystem.

Using Eqs. (5) and (6), we can find the Liouville equation for
$p$-subsystem. For this purpose, let us take a generalized current
vector - $J_p=(\dot{r_k},{\dot{p_k}})$ of the $p$-subsystem in a
phase space [Zaslavsky, 1984]. From Eqs. (5) and (6), we find:
\begin{equation}
{divJ_p=\sum\limits_{k=1}^T({\frac{\partial}{\partial{r_k}}}{V_k}+
\frac{\partial}{\partial{p_k}}{\dot{p_k}})=
\sum\limits_{k=1}^T{\frac{\partial}{\partial{p_k}}{F_{k}^p}}}
\label{eqn7}
\end{equation}

The differential form of particles number conservation law in the
subsystem is a continuity equation:
$\frac{\partial{f_p}}{\partial{t}}+{div(J_pf_p)}=0$, where
$f_p=f_p(r_k,p_k,t)$ is a normalized distribution function of
disks in the $p$-subsystem. With the help of continuity equation
and Eq. (7) for divergence of a generalized current vector in a
phase space, we can show that:
$\frac{df_p}{dt}=\frac{\partial{f_p}}{\partial{t}}+\sum\limits_{k=1}^T({V_k}
\frac{\partial{f_p}}{\partial{r_k}}+\dot{p_k}\frac{\partial{f_p}}
{\partial{p_k}})=\frac{\partial{f_p}}{\partial{t}}+div(J_pf_p)-f_pdivJ_p=
-f_p\sum\limits_{k=1}^T\frac{\partial}{\partial{p_k}}F_{k}^p$.
Thus, we have:
\begin{equation}
{\frac{df_p}{dt}=-f_p\sum\limits_{k=1}^T
\frac{\partial}{\partial{p_k}}F_{k}^p}
\label{eqn8}
\end{equation}
Equation (8) is a Liouville equation for $p$-subsystem. It has a
formal solution:
\begin {center}
{{${f_p=const\cdot{\exp{[-\int{(\sum\limits_{k=1}^T
\frac{\partial}{\partial{p_k}}F_{k}^p)}{dt}]}}}$} (8a)}
\end {center}
From this solution it follows, that the $p$-subsystem will be in a
stationary state when the external forces disappear,
i.e.$\int{(\sum\limits_{k=1}^T\frac{\partial}{\partial{p_k}}F_{k}^p)}dt=0$.
So, the exponent index of the solution of the Eq. (8) determines
the characteristic relaxation time of the system to the
equilibrium state. Hence, the change of the phase volume of the
$p$-subsystem  will last during time of aspiration of the force,
$F_p$, to zero. How this comes about is discussed bellow. It is
possible for all points of the phase space except for "islands",
filled by periodic and quasiperiodic orbits of some Hamiltonian
systems, see [Loskytov {\&} Mihailov, 1990; Zaslavsky, 1999]. If
at the time of preparation, the system is in such an island, it
will return to it periodically. For periodic points we do not have
mixing, and the correlations do not disappear. We can then say
that the probability of return to the initial periodic point is
determined by the probability of preparing the system at its
periodic point.

Let us consider the question: how is the description of selected
subsystems connected to disks system description as a whole.

As, ${\sum\limits_{p=1}^R{\sum\limits_{k=1}^T
F_{k}^p =0}}$, the next equation for the full
system Lagrangian, $L_R$, will be:
\begin{equation}
{\frac{d}{dt}\frac{\partial{L_R}}{\partial{V_k}}-
\frac{\partial{L_R}}{\partial{r_k}}=0}\label{eqn9}
\end{equation}
and the appropriate Liouville equation is
\begin{equation}
{\frac{\partial{f_R}}{\partial{t}}+{V_k}\frac{\partial{f_R}}
{\partial{r_k}}+\dot{p_k}\frac{\partial{f_R}}{\partial{p_k}}=0}
\label{eqn10}
\end{equation}
The function, $f_R$, corresponds to the full system that is
conservative. Therefore, we have: ${\sum\limits_{p=1}^R
divJ_p=0}$. This expression is equivalent to the next equality:
${\frac{d}{dt}(\sum\limits_{p=1}^{R}\ln{f_p})}$ =
$\frac{d}{dt}(\ln{\prod\limits_{p=1}^{R}f_p})$ =
${(\prod\limits_{p=1}^{R}f_p)}^{-1}\frac{d}{dt}(\prod\limits_{p=1}^{R}{f_p})=0$.
So, $\prod\limits_{p=1}^R{f_p}=const$. In equilibrium state we
have $\prod\limits_{p=1}^R{f_p}=f_R$. Because the equality
$\sum\limits_{p=1}^{R}F_p=0$ is fulfilled at all times, we have
the equality, $\prod\limits_{p=1}^R{f_p}=f_R$, as the integral of
motion. This is in agreement with Liouville theorem eabout
conservation of phase space [Landau {\&} Lifshits, 1973].

The simultaneous fulfillment of conditions of phase volume
preservation for full system and validation of time-dependence
solution (b) of Eq. (8) for nonequilibrium subsystems is correct
only when the centre of mass of a system moves on trajectory,
reversible in time. It should take place irrespective of, whether
all subsystems are in equilibrium or not. Therefore, for the
considered $p$-subsystem only those irreversible redistribution of
phase volume and energy that lead to reversibility of motion of
the centre of mass of all system, is possible.

\begin {center}
{3. The mechanism of equilibration}
\end {center}

Let us show that for a hard-disks system, the external force
decreases due to the mixing properties, i.e.
$\int{(\sum\limits_{k=1}^T
\frac{\partial}{\partial{p_k}}F_{k}^p)}dt\rightarrow0$ when
$t\rightarrow\infty$. This means that because of mixing, the
system will go to a stationary state irrespective of where it was
at the initial time (except the periodic points).

The mixing properties of two hard disks was proved by Sinai [1970]
and Zaslavsky [1984]. It was shown that the correlation function
for colliding disks is: $R(t)~exp(n\ln{K})$, where $K=\rho/2$:
$\rho$-length of free run; $n$-the number of collisions. Therefore
the characteristics time of the decay of correlations, $t_d$, for
regular collision disks with unit diameter through a time interval
$\tau$ is determined by equation: $t_d=\tau/(\ln{\rho/2})$. The
condition $K=\rho/2>1$ is satisfied by rare gas. Hence, two hard
disks are mixed. As will be shown bellow, the mixing property is
both, necessary and sufficient for the equilibration. Although in
the situation when we have several disks, the strict mathematical
proof of a mixing property is absent; nevertheless the existence
of the mixing property is usually accepted a priori as well.

For proving the property of aspiration of the resulting force to
zero for a hard-disks system, we shall simplify Eq. (2) for the
$p$-subsystem. Let us assume, that all disks collide
simultaneously in equal, short enough intervals of time, $\tau$.
It is clear, that if for this condition the system goes to
equilibrium then does so definitely for the general case. After
such simplification, Eq. (2) for a p-subsystem, can be written as
\begin{equation}
\dot{V}_k^n={\varphi^n}_{ks(n)}\Delta_{ks(n)}^{n-1}\label{eqn11}
\end{equation}
Here, to each "$k$" disk from $p$-subsystem in moment of time,
$n\tau$ corresponds to "$s(n)$", disk from other subsystems;
$n=1,2,3,...$.

The evolution of the $p$-subsystem is determined by the vector,
$\vec{V}_T^p$, with components denoting the velocities of disks of
the $p$-subsystem: ${\vec{V}_T^p= \{{V_k^p}\}, k=1,2,3,...,T}$.
Some of the time-dependent properties of this subsystem will be
determined by studying the sum of it components. Let us designate
this sum as $\Upsilon_p$. Carrying out the summation in (11) over
all disks of the $p$-subsystem, we obtain:
\begin{equation}
\dot{\Upsilon}_p^n=\sum\limits_{k=1}^T{{\varphi^n}_{ks(n)}\Delta_{ks(n)}^{n-1}}=
F_{P}^n\label{eqn12}
\end{equation}

Equation (12) describes the change of total momentum, acting
on the $p$-subsystem as a result of collisions at time
$n\tau$. The aspiration of a total momentum to zero is equivalent
to aspiration to zero of the force, $F_{P}^n$.

Now let us show, if the mixing property for a disks system is
assumed, the homogeneous distribution of impact parameters of
disks occur as well.

In accordance with the mixing condition, we have the following
[Loskytov {\&} Mihailov, 1990],
\begin{equation}
\mu(\delta)/\mu(d)=\delta/d \label{eqn13}
\end{equation}
Here, $\mu(d)$, is a measure corresponding to the total value of
impact parameter - "$d$"; $\delta$ is an arbitrary interval of the
impact parameter and, $\mu(\delta)$, is a corresponding measure.
Equation (13) implies the number of collisions is proportion to
the interval "$\delta$". It also implies that the distribution of
the impact parameters is homogeneous.

As is well known, see e.g. [Loskytov {\&} Mihailov, 1990;
Zaslavsky, 1984], for mixed systems of correlations decay. For Eq.
(12) this condition can be written down as
$<\varphi_{ks(n)}^n\Delta_{ks(n-1)}^{n-1}>=
<\varphi_{ks(n)}^n><\Delta_{ks(n-1)}^{n-1}>$ i.e. the average from
two multiplied functions is equal to the multiplication of the
average of these functions. $\varphi_{ks(n)}^n$ come from impact
parameters, and $\Delta_{ks(n-1)}^{n-1}$ come from relative
velocities of colliding disks. Therefore this condition is similar
to the condition of independence of coordinates and momenta,
widely used in statistical physics see e.g. [Rumer {\&} Ryvkin,
1977].

Thus, it is possible to carry out the summation in the multiplier,
$\varphi_{ks(n)}^n$, over impact parameters, independent of the
summation of $\Delta_{ks(n-1)}^{n-1}$ over relative velocities of
colliding disks. Then, under the condition of the homogeneous
distribution of impact parameters and when $T>>0$, we can go from
summation to integration. We will have, see [Somsikov, 2001a]:
\begin{equation}
\phi=1/T\lim\limits_{T\rightarrow\infty}\
\sum\limits_{k=1}^T\varphi_{ks}^n=
\frac{1}{G}\int\limits_0^{\pi}\varphi_{ks}^nd(\cos\vartheta)=-\frac{2}{3},
\label{eqn14}
\end{equation}
where $G=2$ is the normalization factor.

Taking into account (14), we have from Eq. (12):
\begin{equation}
{\dot{\Upsilon_p}=-\frac{2}{3}\sum \limits_{k=1}^T\Delta_{ks(n)}}.
\label{eqn15}
\end{equation}
The negative sign in the right-hand side Eq. (15) means, that the
force, $F_p$, decreases.

The next question is about the stability of a stationary point.
The stability of a stationary point of $p$-subsystem can be
established with the help of the Eq. (15). We set the initial
deviation from stationary point, and then consider, how this
deviation changes with time.

Let the point, $Z_0$, be a stationary point, so that,
$F_{P}$, acting on $p$-subsystem, is zero. From the
Lyapunov's theorem about stability it follows that the point, $Z_0$,
is asymptotically stable if any deviation from it gets attenuated.

Let us expand the left and right-hand sides of the equation (15)
in a series by small parameter, $\upsilon$, of perturbation of
velocities of disks of the $p$-subsystem, near point, $Z_0$, and
keep terms up to first-order infinitesimal. The expansion of the
left-hand side of the Eq. (15) gives:
$\dot{\upsilon}=\sum\limits_{k=1}^T\dot{\varepsilon}_k$, where the
summation is carried out on the components of the variation,
$\upsilon$. In the expansion of the right-hand side, there remains
only $(-\frac{2}{3})\sum\limits_{k=1}^T\varepsilon_k=
-2/3\upsilon$. A contribution into the expansion is given by
collisions of disks of the $p$-subsystem, with disks of its
complement. We have:
\begin{equation}
\dot{\upsilon}^n=-\frac{2}{3}\upsilon^{n-1}. \label{eqn16}
\end{equation}

Equation (16) means, that any deviation from an equilibrium state
will decay. Hence, the stationary point in the presence of mixing
is steady. Stability is provided by emergence of returning force,
$F_p$, at a deviation of a subsystem from an equilibrium point. We
shall note, that the Eq. (16) also follows from the theory of
fluctuations see [Landau, 1976].

Let us show, that the emergency of force due to deviation of system
from its equilibrium, provides restriction of spontaneous fluctuations.

Take the system in a nonequilibrium condition. As follows from the
previous statement, any nonequilibrium condition is characterized
by the force, $F_p$, which acts on the $p$-subsystem. This force
is determined by Eq. (12). The time decrease of the force, $F_P$,
is determined by equation: $t_{din}=\int\frac{d\Upsilon}{F_P}$.
Thus, if the system somehow appears at a nonequilibrium point,
over a characteristic time, $t_{din}$, then it should return to
equilibrium.

For the further consideration we shall accept two statements which
follow from the mechanism of equilibration.

First: the degree nonequilibrium is defined by the force, $F_p$.
So, there is a mutual unique conformity between $F_p$ and a phases
space points.

Second: we shall consider, that the spontaneous deviations
increase of system from an equilibrium condition occurs under
increase of $F_p$. The deviation is proportional to force.

If these conditions are fulfilled, it is possible to prove that
such fluctuations are realized only, if $t_{din}>t_{prob}$. Time,
$t_{prob}$ is determined by probabilistic principles. According to
the formula Smoluhovsky [Zaslavsky, 1984], for the case of an
ergodic system, average resetting time, $t$, or Poincare' cycle
time, is equal to $t_{prob}=t(1-P_0)/(P_0-P_1)$, were $P_1$ is the
probability of reversibility during the time $t$, ${P_0}$ is the
probability of initial phase region. Suppose that the system in a
probabilistic way begins to deviate from equilibrium. The
characteristic time of a deviation to any point, $Z_p$, should be:
$\sim{t_{prob}}$. But because in this point the force, $F_p$, acts
on subsystem as returning, the system will approach to the point,
$Z_p$, in a probabilistic way if only $t_{prob}<t_{din}$.

Thus, within the scope of assumption made in this work, the
dynamics of a hard-disks system is completely determined by the
deterministic classical mechanics. Therefore, the need for
probabilistic principles in the description of evolution of the
system and its area of their use are determined by roughness of
transition from summation to integration on impact parameters, and
also from the periodic or quasiperiodic points at the time of
preparation.
\\
\begin {center}
{4. Conclusion}
\end {center}

Splitting conservative, nonequilibrium system as a set of
interacting subsystems, and the analysis of the evolution of
forces of interaction between these subsystems is a basic idea
proposed in this paper. This approved allows us to take into
account the exchange of energy between subsystems and a feedback
between active and inertial forces of their interaction. The
feedback provides relaxation of the system to equilibrium.
Therefore this approach is applicable for studying evolutionary
processes in open systems with polygenic forces and nonholonomic
connections.

The analysis nonequilibrium evolution is based on generalized
equation of Lagrange set for a selected subsystem. The right-hand
side of this equation is a polygenic force of interaction of
subsystems. Because of mixing, this force tends to zero. It causes
evolution of system to equilibrium state. The establishment of
equilibrium is possible starting from all the points of the phase
space except for periodic points. In the vicinity of equilibrium,
where, $F_p$ is near zero, the statistical theory of fluctuations
constructed on the basis of the canonical Hamilton equations is
applicable.

Stability of an equilibrium state is ensured by the emergence of
the returning force when a system deviates from equilibrium. It
imposes appropriate restrictions on amplitudes of probable
fluctuations of system.

Within the framework of this study, the dynamics of the system is
deterministic. Probabilistic principles enter only by the
uncertainty of initial conditions, and coarse graining
of transition from the discrete to the continuous.

Let us compare our explanation of irreversibilities with existing
one, see, for example [Zaslavsky, 1984]. In accordance with this
explanation, irreversibility is a consequence of "coarse graining"
of the phase space. The mixing implies and institutes average
procedure. As a consequence, the information on separate phase
trajectories is lost. This is equivalent to the irreversibility.
The shortcoming of such an explanation is that the nature of
averaging in a phase space because the dynamic equations do not
contain the mechanism of coarse graining.

Though the mechanism offered here is also based on mixing
property, using a force, $F_p$, as evolution parameter, allows
getting rid of explicit use of "coarse graining" idea. In contrast
to directly "coarse graining" of the phase space, the transition
to integration on impact parameters does not deform the nature of
aspiration of the system to an equilibrium state. Here, the role
of replacement of summation on integration on impact parameters
consist in transition from discrete functions to continuous
functions convenient for differential calculus.

\end{document}